\documentclass[
apl,
sd,
amsmath,amssymb,
preprint,
superscriptaddress
]{revtex4-1}

\usepackage{graphicx}
\usepackage{dcolumn}
\usepackage{bm}

\begin{document}


\title{Probing InAs quantum dot states with surface plasmon polaritons}

\author{F. W. de Alencar Sobreira}
\email{fwellysson@gmail.com}
\affiliation{
S\~{a}o Carlos Institute of Physics, University of S\~{a}o Paulo,\\PO Box 369, 13560-970, S\~{a}o Carlos, SP, Brazil
}%

\author{Yu. I. Mazur}
\affiliation{ 
University of Arkansas, Department of Physics, Fayetteville, Arkansas, 72701
}%

\author{Xian Hu}
\affiliation{ 
University of Arkansas, Department of Physics, Fayetteville, Arkansas, 72701
}%

\author{G. J. Salamo}
\affiliation{ 
University of Arkansas, Department of Physics, Fayetteville, Arkansas, 72701
}%

\author{E. Marega Jr.}
\affiliation{
S\~{a}o Carlos Institute of Physics, University of S\~{a}o Paulo,\\PO Box 369, 13560-970, S\~{a}o Carlos, SP, Brazil
}%


\date{\today}

\begin{abstract}
When placed near a thin metallic film, excitons in a quantum dot (QD) decay into surface plasmon polaritons (SPPs), guided modes of light confined at the interface of the metal/dielectric. It was reported that the interaction of SPPs with excitons in a QD may give rise to a modification in the photoluminescence (PL) spectra of the exciton energy levels, however the mechanism that explains this modification remains unclear. We propose a new methodology to excite many energy levels in an exciton by placing the QDs near a metallic array of slits fabricated onto a thin metallic gold (Au) film. By changing the position under the array where the QDs are excited we can resolve between the many energy levels. The unique features of the PL spectra obtained are the result of the strong electromagnetic field (EMF), the increase in the local density of states (LDOS) and the mesoscopic effects attributed to the gradient of the SPP field near the QDs. 
\end{abstract}

\maketitle


The area of plasmonics has grown in the last decade as a promising successor for electronics and photonics, because we can in principle design devices as fast as in photonics and also as small as in electronics\cite{Maier2001, Ozbay2006, Atwater2007}. To do so, the interaction of SPPs with matter has become of fundamental importance, even more because of the ability to control quantum emitters at low dimensional systems is a natural feature of plasmonics\cite{Barnes2003, Chang2006, Chang2007, Chang2009, Gramotnev2010}. For example, it has been reported that when SPPs interact with semiconductor quantum emitters such as excitons in QDs or quantum wells, they may give rise to an enhancement or quench in the PL spectrum of these emitters\cite{Khurgin2009, Kulakovich2002, Khatab2013, Yuan2013}, although details about the conditions for each case are still lacking.

Another important characteristic of SPP-exciton interaction is the ability to manipulate the spectrum of these quantum emitters, as the SPP EMF interacts differently with each energy level than does the free propagating light in a semiconductor medium. This is important since the total intensity of PL spectrum is important for the design of electronic devices such as lasers and optical switches. It was also shown that a strong coupling regime of SPPs with quantum emitters placed near a plasmonic structure can be achieved\cite{Chang2007a, Gonzalez-Tudela2013}. This strong coupling is a result of the change in the local density of states of light near the emitters and can give rise to new exotic states such as the so-called quantum fluids of light\cite{Carusotto2013}, which is a promising subject to understand the fundamental properties of light. In order to investigate this and other properties of this interaction many efforts have been made to probe SPP interaction with matter \cite{Fedutik2007, Vasa2008, Andersen2010, Passmore2011}. However, we take a different approach that employs a metallic array in which SPPs can be excited at different positions and its interaction with excitons in QDs are probed using a microPL system.


The sample was grown by Molecular Beam Epitaxy on (100)-oriented GaAs substrate. After the desorption of the GaAs oxide, a $300$nm buffer layer was grown at a substrate temperature of $580^{\circ}{\rm C}$. We used reflection high-energy electron diffraction to observe a streaky ($2\times 4$) reconstruction at the end of the buffer layer growth. After the buffer we deposited $2.0$ monolayers (ML) of InAs at $530^{\circ}{\rm C}$ followed by a $30$s annealing. Then a $12$nm GaAs layer was grown. This process was repeated 5 times in order to produce a sample of stacked QDs to increase the density of QD emitters. The $12$nm spacer layer is enough to prohibit electronic coupling between the QDs at different layers.

\begin{figure}
\includegraphics{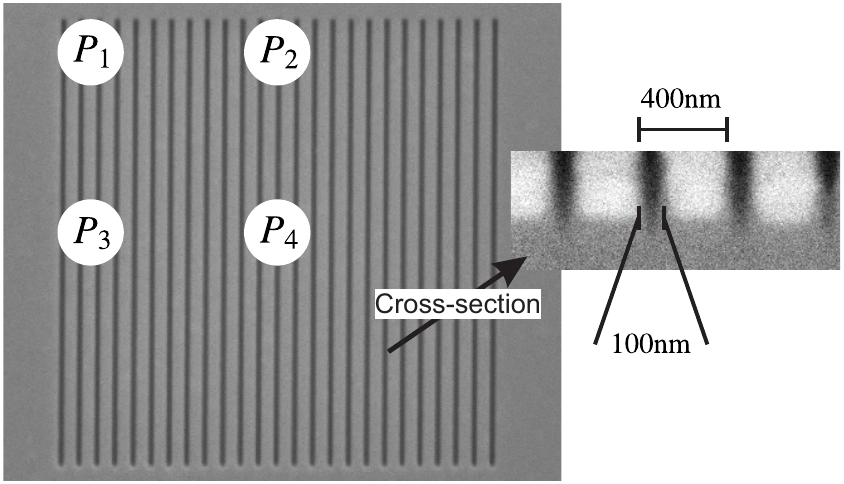}
\caption{\label{fig:FIBpattern} Secondary electron microscope image of the slit array patterned using FIB on a Au film of 120nm thickness deposited over a GaAs cap layer. The positions numbered $P_{1-4}$ represent the positions where the PL spectrum of the sample was taken with laser focused on the surface of the sample. The inset shows a cross-section image of the pattern. The slits are $100$nm width, $10\mu$m length and the period of the array is $400$nm.}
\end{figure}

After the QD growth a thin Au film of $120$nm was deposited over the final GaAs surface by thermal evaporation. Then, we used a Focused Ion Beam (FIB) FEI Quanta 3D 200i system to mill a pattern over the Au film using a Ga$^+$ ion beam. The pattern consists of a slit array with $100$nm width and period (distance between the center of two adjacent slits) of $400$nm, as shown in fig. \ref{fig:FIBpattern}.

As the patterning of the array is processed, the semiconductor underneath the slit becomes amorphous and is partially sputtered, as shown in the inset of fig. \ref{fig:FIBpattern}. After this sputtering all the QDs up to a distance of around $30$nm below the slit were destroyed. As the ion beam penetration is mostly perpendicular to the (100) direction, the QDs below the metallic film are protected and their quantum efficiency is not affected. The dielectric constant of the semiconductor is not very sensitive to these damages, then the spectra of the SPPs generated is not affected by the amorphization process.

We measured the PL spectra of the sample on different positions, for the full pattern and also for the reference sample without the Au film (UQDs) at nitrogen temperature ($77$K) using a Horiba-Jobim Yvon LabRAM microPL system equipped with an infinite focal length distance lens. The sample was excited using a $633$nm HeNe laser with $3.8$mW and a $785$nm Ti-Sapphire laser of $16.5$mW, maximum power measured at sample position. The electric field of the laser beam could be polarized parallel (TE mode) or perpendicular (TM mode) to the direction of the slits in the array. The waist of the laser beam used is $2\mu$m at the focus, 5 times smaller than the length of the slits used in the patterns. With the sample on the laser focus we measured the PL spectrum at four independent positions, see fig. \ref{fig:FIBpattern}.

\begin{figure}
\includegraphics{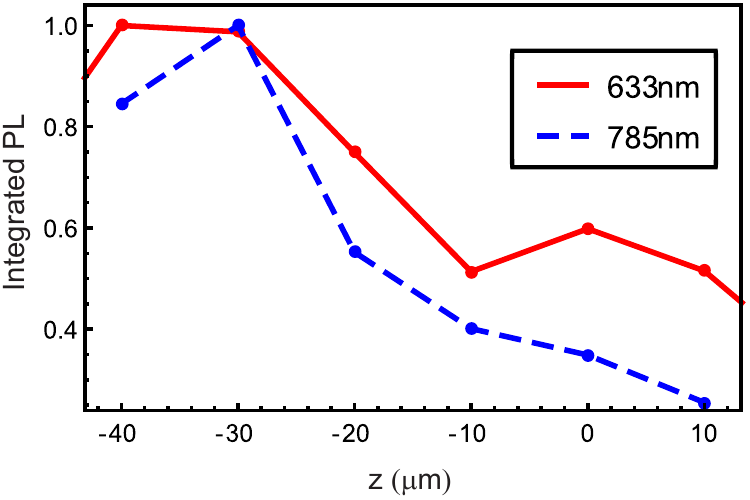}
\caption{\label{fig:PL_focusdistance} Integrated PL signal for the plasmonic array for $633$nm (red line) and $785$nm (blue dashed) lasers. The maximum signal was normalized to unity for both lasers, and is reached when the plasmonic array is between $30\mu$m and $40\mu$m below the laser focus, the position $z = 0\mu$m corresponds to the sample on the laser focus. The experimental points are connected with lines as a guide to the eye. See fig. \ref{fig:TE_TM_mode} for the PL signal.}
\end{figure}

We defocused the laser beam, with its center placed at point $P_4$, in order to illuminate the full pattern with almost the same intensity at every point, then we are able to investigate the effect of the finite dimensions of the plasmonic array. As we place the sample below the laser focus the integrated PL signal reaches a maximum, indicating that more QDs are excited by the SPPs and vice-versa, and then decreases, as the total power incident over the array decreases. This result is shown in fig. \ref{fig:PL_focusdistance} for both $633$nm and $785$nm lasers using TM polarization. The position where the maximum signal for each laser is obtained slightly varies because of the dependence in the transmission over the array with the incident wavelength, however for both lasers it is between $30\mu$m and $40\mu$m over the sample.

\begin{figure}
\includegraphics{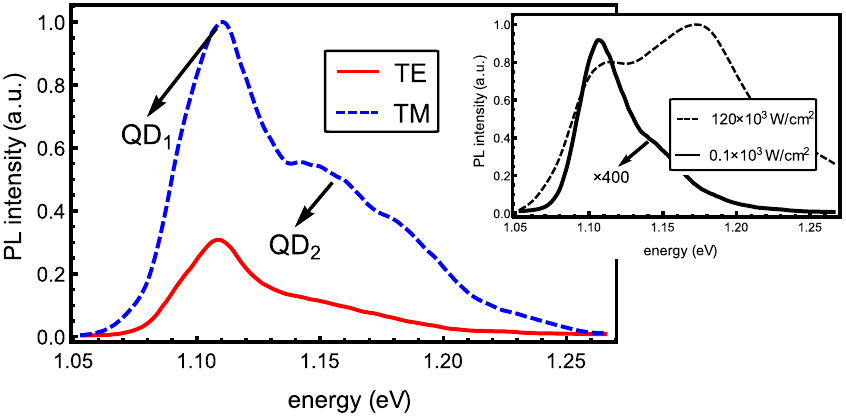}
\caption{\label{fig:TE_TM_mode} PL signal for TE mode (red line) and TM mode (blue dashed) for the $633$nm laser incident on the plasmonic array. The inset shows the PL signal for the UQDs at low ($0.1\times 10^3$ W/cm$^2$) and high ($120\times 10^3$ W/cm$^2$) intensity of the laser field.}
\end{figure}

The first step to demonstrate that boundary effects are relevant to understand the SPP modes supported by the array and their interaction with the energy levels of the QDs, we change the laser polarization and analyse the QD PL spectra for both TE and TM laser polarization (see fig. \ref{fig:TE_TM_mode}). Fig. \ref{fig:TE_TM_mode} shows the PL spectra for laser incidence at the focus position which maximizes the integrated signal using the $633$nm laser.

In the ideal case of an infinite slit array, no signal is expected for the TE mode because there is no SPP excitation for this polarization, so the electric field over the QDs should be zero. However, as each slit in the array has a finite length it is possible to excite SPPs even using TE polarization, because the incident light couples with the SPPs in the edges of each slit. Also, the integrated PL signal shows that although there is a non vanishing signal for TE mode this signal is still smaller when compared to TM mode.

This result is consistent with the expected value for light transmission calculated using a Finite Difference Time Domain (FDTD) simulation which is $20$\% ($45$\%) for the TM mode and $0.7$\% ($0.4$\%) for the TE mode using a $633$nm ($785$nm) laser excitation. In the simulation the array is excited by a plane wave (slit array far from laser focus) and the semiconductor is filled with GaAs only, without the InAs layers. In the simulation, the EMF due to laser decreases by less than $5\%$ between the top and the bottom layers of the stack of QDs.

As shown in the inset of fig. \ref{fig:TE_TM_mode} there are peaks at two different energies. These peaks correspond to different families of QDs, QD$_1$ at $1.11$eV and QD$_2$ at $1.16$eV because they are present even at low power excitation in the UQDs sample.

\begin{figure}
\includegraphics{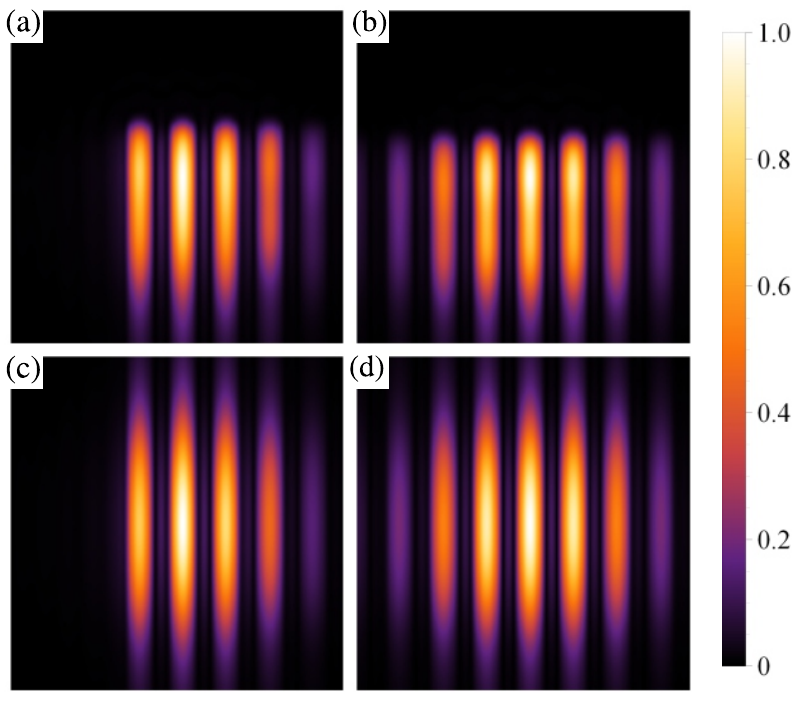}
\caption{\label{fig:intensity_FDTD} FDTD simulation for the electric field intensity profile of the SPP excitation over the sample of stacked QDs. The result for the points $P_1 - P_4$ given by fig. \ref{fig:FIBpattern} are shown from (a) to (d), respectively. In the simulation we used a $785$nm laser with gaussian beam profile at TM mode and beam waist of $2\mu$m as used in the experimental setup, and the profiles were taken 12nm below the metal/semiconductor interface (the position of the nearest layer of QDs).}
\end{figure}

In order to investigate the response of the system to the intensity of the SPPs we have focused the laser beam at the four different points shown in fig. \ref{fig:FIBpattern}. We used a $785$nm laser beam at full power and TM polarizations. Using an FDTD model of a gaussian beam with $2\mu$m beam waist incident on each of these four points we were able to obtain the intensity profile of light over the sample of QDs. Fig. \ref{fig:intensity_FDTD} shows the electric field intensity at the layer of QDs placed nearest the metallic film.

\begin{figure}
\includegraphics{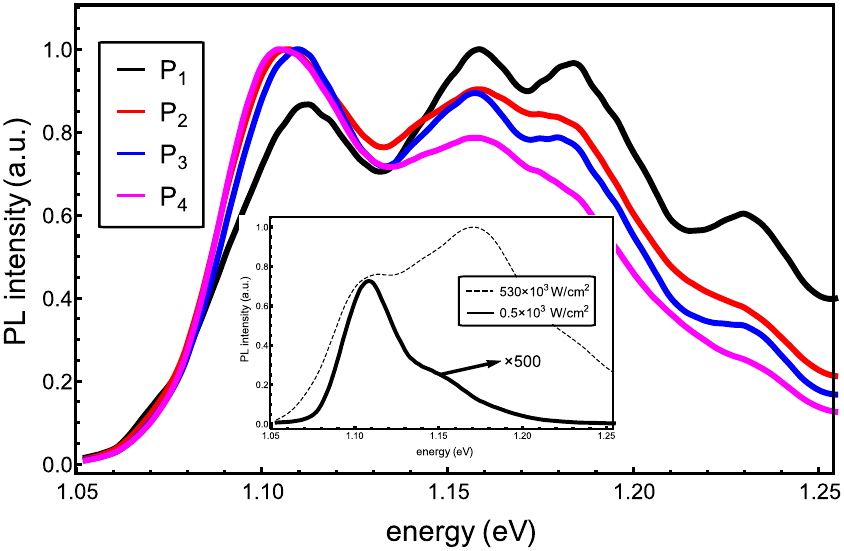}
\caption{\label{fig:PL_1234} Normalized PL signal for each of the 4 points $(P_{1-4})$ given by fig. \ref{fig:FIBpattern}. The $785$nm laser beam is focused at each point at maximum intensity of $530\times 10^3$ W/cm$^2$ and TM polarization mode. The inset shows the normalized PL spectra for the sample of UQDs at minimum ($0.5\times 10^3$ W/cm$^2$) and maximum ($530\times 10^3$ W/cm$^2$) laser power.}
\end{figure}

The maximum EMF intensity was achieved for position $P_1$ followed by positions $P_2$, $P_3$ and $P_4$, the maximum value of the EMF for these points is given by $1 : 0.88 : 0.95 : 0.78$, respectively. The PL signal for the sample of QDs was measured for each of these points. These results are plotted in fig. \ref{fig:PL_1234}. Our interest is mostly on the shape of the PL signal than for the total integrated PL, for this reason the maximum intensity for each result was normalized to unity. The total integrated PL slightly changes for each point because the total power transmitted by the SPPs depends on where they are excited.

The maximum intensity achieved for each point is an indirect measure of the local density of states of SPPs for each position. To understand why the intensity of light changes for each point where the excitation source is placed we note that the mechanism for SPP excitation is a result of a break in the translation symmetry introduced by the slit array in the direction of electric field polarization. As a result, the TM mode is the most important during the excitation process, that explains why the points 1 and 3 have a stronger intensity.

At point 4 the array behaves as an ideal infinity slit array because its length is much bigger than the beam waist. Although the TE mode coupling to light is not very strong, as shown in fig. \ref{fig:TE_TM_mode}, the edges at the of slits in point 2 slightly enhances the light confinement which results in an increase of the EMF of the SPP when compared to point 4.

As a result of the high intensity of the EMF due to the SPPs it is possible to observe the QDs excited states in the PL at each point. The inset in fig. \ref{fig:PL_1234} shows the PL spectra for the UQDs at the focus for the $785$nm laser beam, measured both at maximum and minimum laser intensity. In the inset we observe that the peak on $1.18$eV, which corresponds to the first excited state of QD$_1$, mixes with the $1.16$eV peak, which is due to the ground state of QD$_2$.

The spectra for points $P_2$ and $P_4$ resembles that for the UQDs with the mixing of the ground and excited states of QD$_2$ and QD$_1$ at $1.16$eV and $1.18$eV, respectively. However, the ground state of QD$_1$ and QD$_2$ is not fully populated (as in the UQDs at low intensity), for this reason there is more intensity at $1.16$eV than in $1.18$eV and no excited state for QD$_2$ is observed.

At point $P_3$, the mixing at $1.18$eV can be well distinguished of that in $1.16$eV and the new peak appearing at $1.23$eV corresponds to the excited state of QD$_2$. At point $P_1$ we observe that the ground state of QD$_1$ becomes fully populated because its intensity has a quench relative to the excited state. At this position, the resolution of the peaks at $1.16$eV and $1.18$eV also increases, and the same happens for the peak at $1.23$eV due to the excited state of QD$_2$.

Then the edges in the plasmonic structure imply in a higher intensity for the EMF, which explain the excitation of the many energy levels in the QD, observed by the saturation of the ground state and the appearing of the excited states.

\begin{figure}
\includegraphics{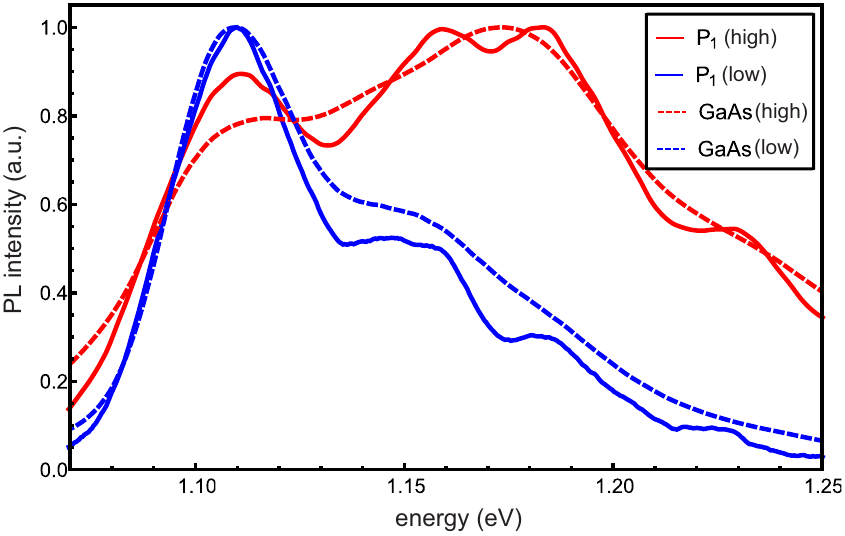}
\caption{PL spectrum for the sample of UQDs (dashed line) and when the laser beam is focused on position $P_1$ (line). The spectrum was taken for laser at maximum intensity (high, $120\times 10^3$ W/cm$^2$) and at two orders of magnitude smaller (low, $1,2\times 10^3$ W/cm$^2$)}
\label{fig:resolution}
\end{figure}

The remarkable resolution of the many energy levels excited in our sample can be explained by a set of unique features of the interaction with SPPs. When compared to the PL spectrum of the UQDs, the spectrum taken for position $P_1$ resolves the many energy levels even at a lower laser intensity, as shown in fig. \ref{fig:resolution}.

This is a result of the fast recombination rate of the exciton energy states to SPPs when compared to light or phonons in the GaAs lattice. The enhancement in the LDOS as we approach a surface plasmon resonance \cite{Schiff2011}, the high value of the coupling factor of the SPP-exciton interaction for energy levels near the QD energy levels \cite{Gonzalez-Tudela2013}, as well as the mesoscopic effects of the exciton wavefunction introduced by the proximity with the metallic film \cite{Andersen2010} are the important mechanisms. The last is more pronounced for excited states where quadrupole moments are more intense \cite{Grundmann1996}.


In conclusion, we observed the coupling between SPP and exciton by the transfer of energy from the exciton energy states and that this interaction is stronger for excited states. The ability to excite and resolve many energy levels in a sample of QDs by only choosing the position over the plasmonic structure where they are excited is an important new feature of SPP interaction which can be used with methods of cavity quantum electrodynamics near the visible range to transfer the emitter states to a plasmonic state \cite{Sørensen2004}. This scheme could also be used to design optoelectronic devices such as optical switches and logical gates for application in quantum information as well as for the investigation of the properties of excited states for an exciton in a QD.


The authors thank Prof. R. G. Pereira for fruitful discussions. We thank the computational support of the N\'{u}cleo de Apoio \`{a} \'{O}ptica e Fot\^{o}nica (NAPOF-USP), as well as technical support of NanoPlasmonics group. We are also grateful to the Brazilian agencies CNPq under grant 
number $140152/2011-3$, FAFQ, FAPESP and CAPES for financial support.


\bibliography{bibfile}

\end{document}